*Original Article*

# Influence of Stakeholder Involvement in M&E on the Performance of Donor-Funded Projects in Informal Settlements in Kisumu Central Sub-County, Kisumu County, Kenya

[1]Brighton Savayi Amuni, [2]Stella Silas Karimi, [3]Felistus Mwikali Patrick
[1,2,3]*The Catholic University of Eastern Africa, Nairobi, Kenya.*

***Abstract:*** *Stakeholder engagement and participatory approaches influence the effectiveness of donor-funded projects. Participatory Monitoring and Evaluation (PM&E) methodologies ensure that local communities play an active role in decision-making, leading to more sustainable outcomes. Given the complex socio-political landscape of Kisumu Central Sub-County, there is a critical need for inclusive and context-responsive project monitoring strategies. Initiatives that have integrated local leaders, youth groups, and women-led organizations into their M&E processes tend to achieve stronger community buy-in, improved continuity, and more impactful outcomes. The current study explored the influence of stakeholder involvement in M&E on the performance of donor-funded projects in informal settlements in Kisumu Central Sub-County, Kenya. The study was guided by Stakeholder Engagement Theory. The study used a Convergent Parallel design with a sample size of 364 respondents computed using Yamanes' Sampling formula ($n = N/(1 + N(e)^2)$), drawn from 27 donor-funded projects in Obunga and Nyalenda informal Settlements. Purposive sampling was used for project managers, project M&E staff, and community members served, while community members were selected using stratified random sampling. The study findings revealed that there were regular opportunities for stakeholder interaction in the projects ($\bar{x}$ =4.05, SD 1.08), stakeholders contributed to the development of the organization/project ($\bar{x}$= 3.79, SD=.940), although stakeholders' perspectives and opinions were not diligently incorporated into programming ($\bar{x}$=2.06, SD=.879) as anticipated. The study, therefore, concluded that stakeholder involvement in M&E influenced the performance of donor-funded projects. The study recommended that the project managers in donor-funded projects need to enhance stakeholder involvement for project ownership and sustainability.*

***Keywords:*** *Donor-funded Projects, Monitoring and Evaluation, Performance, Stakeholder, Stakeholder Involvement.*

## I. INTRODUCTION

According to FAO (2025), components of a Monitoring and Evaluation (M&E), when collaboratively developed with all key stakeholders, promote participation and enhance project/plan ownership. Stakeholder engagement and participatory approaches also influence the effectiveness of donor-funded projects. As posited by Estrella and Gaventa (2019), participatory monitoring and evaluation (PM&E) methodologies ensure that local communities play an active role in decision-making, leading to more sustainable outcomes. Similarly, Rossi, Lipsey, and Freeman (2021) assert that the integration of qualitative and quantitative assessment methods improves the reliability of project evaluations and allows for a more comprehensive understanding of project performance.

Muwanga and Kule (2020) highlight that project outcomes are substantially improved when local stakeholders are actively involved in M&E processes, especially in urban poverty interventions. In their study to determine the factors influencing performance of NG-CDF funded projects, Musomba, Kerongo, and Mutua (2020), there was a statistically significant association between performance and Project Management Skills. The Pearson chi-square for Project Performance and Community Participation implied that there is a statistically significant association between Performance and Community Participation. Over 60% of M&E practices in sub-Saharan Africa exclude beneficiary feedback mechanisms, particularly in urban informal settlements.

In Uganda, the inclusion of community members in monitoring urban water supply initiatives has been associated with noticeable improvements in service delivery and user satisfaction (Namara & Ssozi, 2016). Comparable outcomes have been observed in Tanzania, where the active participation of local health workers in tracking donor-supported health programs contributed to greater service uptake and patient retention (Kusek & Rist, 2020). Such regional experiences illustrate the value of participatory monitoring frameworks in enhancing the effectiveness of development efforts, especially in underserved urban areas.

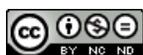

Kenya has long benefited from donor-funded interventions aimed at addressing the multidimensional challenges of urban informal settlements. However, evidence suggests that despite significant investment, the performance and sustainability of these projects are often undermined by insufficient community involvement in monitoring and evaluation processes (Githinji & Were, 2020). In Obunga and Nyalenda informal settlements, where over 100 donor-funded projects operate, inefficiencies such as financial misappropriation, project redundancy, and low community participation have been reported. These shortcomings are frequently linked to weak M&E frameworks and limited stakeholder engagement (Kisumu County Government, 2023). Otieno and Otieno (2019) note that while implementing agencies often possess adequate technical capacity for M&E, they rarely prioritize the systematic involvement of community stakeholders in tracking project progress.

Given the complex socio-political landscape of Kisumu Central Sub-County, there is a critical need for inclusive and context-responsive project monitoring strategies. Initiatives that have integrated local leaders, youth groups, and women-led organizations into their M&E processes tend to achieve stronger community buy-in, improved continuity, and more impactful outcomes (Wachira & James, 2021). The current study, therefore, explores the influence of stakeholder involvement in M&E on the performance of donor-funded projects in informal settlements in Kisumu Central Sub-County, Kenya.

## II. LITERATURE REVIEW

### A) Theoretical Review: Stakeholder Engagement Theory

The origin of Stakeholder Theory can be attributed to the body of literature on management. Preston (1999) contends that the origin of the stakeholders' idea may be traced back to the era of the Great Depression in the United States (1929-1941). At this time, the larger Electric Company recognized four main stakeholder groups: shareholders, employees, consumers, and the wider public. Freeman (1984) asserts that the research conducted by the Stanford Research Institute in 1963 is considered to be the basis of stakeholder theory. According to Freeman (1984), stakeholders are defined as the groups whose support is crucial for the continued survival of the organization, as stated by the institute's definition.

Freeman (1984) emphasizes the importance of managers recognizing four key stakeholders within an organization: firm owners, consumers, employees, and suppliers. The goals of corporate owners underwent a significant shift in the second half of the 20th century. In addition to focusing on monetary gains, they expanded their objectives to include "shareholder activism" and the advancement of social justice. A more realistic depiction of the connections between the various groups of people interacting with and surrounding the business is provided by the stakeholder model framework.

Freeman created a novel and concise conceptual framework to comprehend the structure and functioning of enterprises. This framework incorporates a thorough examination of the scholarly literature on corporate strategy and organisational theory, in addition to substantial research and real-world data. The stakeholder model, first introduced by Freeman (1984), portrays the organization as the central focal point of a wheel, with stakeholders located at the outer ends of the spokes encircling the periphery (Freeman, 1999). The descriptive approach, which seeks to offer a thorough explanation of organisational traits and behaviour, including the management strategies used by corporations and the considerations made by the board of directors towards corporate stakeholders, is one of three different approaches that make up supplier theory. The instrumental approach uses empirical data to show how stakeholder management and organisational goal achievement are related.

There are three different approaches to stakeholder theory. The first is the descriptive approach, which seeks to give a thorough explanation of organisational traits and behaviour, including corporate management techniques and board of directors considerations for corporate stakeholders. To demonstrate the links between stakeholder management and achieving organisational goals, the instrumental approach uses empirical data.

The normative approach entails examining the objectives and purpose of businesses in order to develop moral and philosophical standards that ought to direct their management and behaviour. Two essential principles are included in the above-described concept: According to the Principle of Corporate Legitimacy, a company's management should prioritise the interests and concerns of its stakeholders. It is necessary for stakeholders to actively participate in decision-making processes that have a substantial influence on their general well-being. Ozturk (2012) argues that the stakeholder fiduciary concept requires managers to serve as advocates for the stakeholders in order to safeguard the corporation's interests and ensure its long-term viability.

The theory used in this study emphasises the importance of stakeholder involvement in programming as well as the several approaches that may be used to guarantee the best possible performance from the organisation. However, the theory did not explain how this optimal performance should be achieved by maintaining a balance in organisational resources and resource priorities. This calls for the implementation of the Resource-based Theory, which revealed that, in order to achieve the best project performance, organisational resources and resource priorities must be balanced.

*B) Empirical Review: Stakeholder Involvement and the Performance of Donor-Funded Projects*

Participatory project planning is a collaborative approach that involves stakeholders in the design and decision-making processes of sponsor-funded projects. Incorporating the viewpoints and needs of recipients ensures enhanced effectiveness and enduring outcomes (Panday, Wang, & Spasova, 2023). Engaging stakeholders from the outset via participatory project planning improves openness, accountability, and ownership. This inclusive technique enhances project performance by fostering shared accountability, augmenting local skills and ideas, and promoting communication and cooperation. Participatory project planning in Kisumu's donor-funded projects promotes the successful achievement of development objectives by empowering local people and augmenting their potential for sustained success.

Panday, Wang, and Spasova (2023) discovered that the use of participatory project planning in Khulna markedly enhanced community engagement and involvement. Participation in project planning in Khulna promotes the development of relevant and effective projects. The research established that the participatory strategy ensures the involvement of all stakeholders, including community people, local authorities, NGOs, and funders, in the decision-making process. This research was done independently in Khulna, India. Therefore, the data and conclusions may lack importance or reliability in the context of Kenya. Consequently, this inquiry is imperative.

Panday & Panday (2020) aimed to determine the factors affecting the effectiveness of slum upgrading projects and to analyze the influence of stakeholder participation at various stages of the project life cycle on its success. This study used a descriptive research approach. According to the results, a participative approach increases the likelihood that slum redevelopment initiatives will be effective. This study was conducted exclusively in Mumbai, India. Therefore, the findings might not apply to Kisumu's informal settlements. The purpose of this study is to investigate how stakeholder participation affects the effectiveness of donor-funded initiatives in Kisumu, Kenya.

Pereira et al. (2021) conducted research to emphasize the need for a reevaluation of the current wheat policy and practices in Ukraine. They promoted the use of collaborative and decentralized planning approaches. The study took a descriptive cross-sectional method and applied project management ideas. According to the research findings, performance may be enhanced by including iterative combinations of planning criteria from the project management life cycle into participatory project planning. The results of this study are limited in their generalisability because it was conducted mostly in Ukraine and employed a cross-sectional methodology. This study examined donor-funded initiatives in Kisumu, Kenya, using a convergent parallel design.

Tumwebaze and Irechukwu (2022) investigated the importance of stakeholder engagement in the skills development initiative in Rwanda's Gasabo District. A descriptive research approach was used to survey 211 persons from a total sample size of 400. The study findings indicated that stakeholder involvement enhances the durability of skill development projects. A significant and positive association was observed between stakeholder impact and KPIs, namely beneficiary satisfaction ($r = 0.928$) and beneficiary ownership ($r = 0.94$). This research was only done in Rwanda; hence, the results cannot be generalized to Kenya. Consequently, the present investigation is necessary.

Emmanuel (2020) did a study in Rwanda to examine the influence of stakeholder engagement on the performance results of donor-funded projects. The study used a descriptive design, with a sample of 75 participants. The data unequivocally indicated that the efficacy of the EYICM project in Bugesera District was significantly influenced by the active participation of stakeholders. This research largely examined the effects of stakeholder participation. This research examines several methods for stakeholder participation in programming, which might significantly enhance project results in donor-funded programs in Kisumu.

In Tharaka Nithi, Gilbert and James (2021) conducted research to examine the impact of stakeholder participation, financial variables, monitoring and evaluation techniques, and technology elements on the successful implementation of donor-funded projects. The research used a descriptive technique to analyse 34 donor-funded programs presently implemented in the county. The study revealed that stakeholder participation, financial support, monitoring and assessment, and technological advancements favourably impacted the execution of donor-funded projects. Nonetheless, after conducting a thorough analysis of several characteristics, the research could not conclusively determine the importance of stakeholder engagement. This study significantly enhances prior research by examining the subject of stakeholder engagement in more detail. It especially highlights a selection of designated indicators: communication, reporting, decision-making, and interaction frequency.

*C) Research Methodology*
  a. **Research Design**

The study used a Convergent Parallel design, which allowed for the concurrent collection and analysis of qualitative and quantitative data, promoting a thorough comprehension of the investigated issue (Creswell, 2014). The study was conducted in Nyalenda and Obunga, two informal settlements located within Kisumu Central Sub-County, with 27 humanitarian donor-funded projects. The study target population was 4050 drawn from the 27 humanitarian project staff and the served community.

b. **Sample Size**

The sample size was computed using Yamane's (1967) sampling formula, stated as: $n = N/(1 + N(e)^2$

Whereby: n = sample size, e = level of confidence (at 95%, e=0.05) and N = Total population

Therefore, the sample size of the served community members was computed as follows:

$$n = N/(1 + N(e)^2$$
$$n = 4050/(1 + 4050(0.05)^2$$
$$n = 364.04$$
$$Therefore, n \approx 364\ respondents$$

The sample size for the study was therefore 364 respondents (served community members) with 108 purposively sampled project staff and community leadership.

**Table 1: Sample size**

| Respondents' categories | sample size | Sampling technique |
|---|---|---|
| M&E Officers | 27 | Purposive |
| Project Managers | 27 | Purposive |
| Field Officers | 27 | Purposive |
| Community leaders | 27 | Purposive |
| Community members served | 364 | Stratified Random |
| **Totals** | **472** | |

*Source:* Author, 2025

c. **Sample Strategy**

A multistage sampling strategy was used in the study to choose project participants. The researcher deliberately chose two unofficial communities that were a part of donor-funded projects during stage one. A census sampling of all 27 donor-funded programs operating in these two informal settlements was also carried out by the researcher. Project managers, M&E officers, and community leaders were specifically selected for stage two of the study based on their possession of particular data. To choose the community members serviced, stratified random selection was employed. Here, a list of community members was gathered and arranged by informal settlement and the project they are associated with. After that, they were chosen at random. Members of the community were chosen at random from the two target informal settlements; Nyalenda had more responders because of its location and population, as well as the fact that it had more donor-funded projects running there.

## 3.6 Data Collection Instruments

In this study, data were primarily gathered through interviews, using researcher-administered questionnaires and interview guides designed for key informants. Data was collected from 27 M&E officers, 27 field officers and 364 community members served through the administration of questionnaires by the researcher. The surveys consisted of carefully designed questions that aimed to facilitate the collection of quantitative data. The Key Informant Interviews (KIIs) were utilized to gather qualitative data from the 27 project managers and 27 community leaders.

## III. RESULTS AND DISCUSSION

The literature makes clear that project stakeholders have made a substantial contribution to the expansion of initiatives. The goal of the study was to set the scene for donor-funded initiatives in Kisumu's informal settlements. The findings on a number of indicators chosen to evaluate stakeholder involvement are thus presented in this section. These indicators include the types of stakeholders participating, the frequency of involvement, the ways in which stakeholders were involved, and the frequency of projects hosting forums for stakeholder engagement.

The study sought to establish the various stakeholders that were involved in donor-funded projects. The findings are as illustrated in Table 2. From the findings, Area local administrators, project staff, and project beneficiaries were the most involved stakeholders in programming, as indicated by 14.4% (44) of the participants who took part in this study in each category, who were on the affirmative.

**Table 2: Project stakeholders**

| Stakeholders | Frequency | Percent |
|---|---|---|
| Donors/ funding partners | 33 | 10.7 |
| Surrounding Community members | 44 | 13.8 |

| Area local administrators | 44 | 14.4 |
|---|---|---|
| Project staff | 44 | 14.4 |
| Project beneficiaries | 44 | 14.4 |
| County/ national governments | 40 | 12.9 |
| Religious leaders | 30 | 9.7 |
| Others | 30 | 9.7 |
| **Total** | **309** | **100.0** |

***Source:*** *Field Data (2025)*

According to 13.8% (44) and 12.9% (40) of the participants in this study, the government and the surrounding communities were also deemed important in the work of projects. Because they were frequently only updated on the projects' status at predetermined reporting intervals after funding them, the project donors/funding collaborators, spiritual leaders, and other stakeholders were not equally involved as the others. This was indicated by 10.7% (33), 9.7% (30), and 9.7% (33), respectively. These findings generally showed that programs welcomed stakeholder input and made every effort to include them in programming. Due to their operations in informal settlements, some projects did this for convenience in terms of both their own and the project equipment's safety.

**Table 3: Ways in which stakeholders were involved in projects**

| How Stakeholders were involved | Frequency | Percent |
|---|---|---|
| Needs assessment | 44 | 14.3 |
| Project Design | 12 | 3.9 |
| Baseline assessment | 51 | 16.4 |
| Project evaluation | 39 | 12.6 |
| Project monitoring | 53 | 17.3 |
| Meetings and workshops | 55 | 17.8 |
| Outreach activities | 55 | 17.8 |
| **Total** | **309** | **100.0** |

***Source:*** *Field Data (2025)*

Stakeholders participated in programming at various levels throughout the project life cycle and in diverse ways. The study aimed to determine the extent of stakeholder involvement in the donor-funded initiatives examined. All 309 participants in the study provided a response to this question. The results are presented in Table 3. Stakeholders predominantly participated in meetings, workshops, and community outreach initiatives. Seventeen point eight (17.8%) (55) of the participants in this survey indicated this, respectively. The respondents said that the project beneficiaries exhibited greater engagement in outreach activities conducted by the donor-funded initiatives. All additional stakeholders and project beneficiaries will be invited periodically to meetings and workshops as necessary.

Stakeholders were increasingly engaged in project monitoring, baseline assessments, and needs assessments, as evidenced by 17.3% (53), 16.4% (51), and 14.3% (44) of the research participants, respectively. The project design was predominantly seen as the exclusive responsibility of the project personnel, resulting in minimal engagement of other stakeholders, as evidenced by 3.9% (12) of the participants in this survey. Consequently, several projects were conceived based on the perceived needs of the communities rather than their actual felt needs, suggesting that some initiatives prioritised issues of greater significance to the community over those being executed.

**Table 4: Stakeholder involvement**

| Statement | SD (F, %) | D (F, %) | N (F, %) | A (F, %) | SA (F, %) | Mean | Std. Dev. |
|---|---|---|---|---|---|---|---|
| The voice of the project stakeholders counts in the design of project activities/ initiatives | 48 (15.5%) | 32 (10.4%) | 24 (7.8%) | 159 (51.5%) | 46 (14.9%) | 3.40 | 1.30 |
| There is frequent communication with stakeholders on project implementation | 16 (5.2%) | 24 (7.8%) | 136 (44.0%) | 95 (30.7%) | 38 (12.3%) | 3.37 | 0.97 |
| There is a clear communication and feedback structure in the organization | 64 (20.7%) | 24 (7.8%) | 48 (15.5%) | 135 (43.7%) | 38 (12.3%) | 3.19 | 1.34 |
| Stakeholders contribute to the growth of this organization | 16 (5.2%) | 24 (7.8%) | 8 (2.6%) | 223 (72.2%) | 38 (12.3%) | 3.79 | 0.94 |
| There are frequent forums where stakeholders are engaged in our work | 8 (2.6%) | 32 (10.4%) | 24 (7.8%) | 119 (38.5%) | 126 (40.8%) | 4.05 | 1.07 |
| The organization has not conducted stakeholder mapping and has an exhaustive list of stakeholders | 48 (15.5%) | 139 (45.0%) | 90 (29.1%) | 16 (5.2%) | 16 (5.2%) | 2.39 | 0.98 |

| The project constantly reports on project progress to the project stakeholders | 48 (15.5%) | 24 (7.8%) | 16 (5.2%) | 147 (47.6%) | 74 (23.9%) | 3.57 | 1.35 |
|---|---|---|---|---|---|---|---|
| The community members served are not part of the key project stakeholders | 8 (2.6%) | 200 (64.7%) | 26 (8.4%) | 55 (17.8%) | 20 (6.5%) | 2.61 | 1.02 |
| The views and opinions of stakeholders are not taken into account in our programming | 63 (20.4%) | 206 (66.7%) | 8 (2.6%) | 24 (7.8%) | 8 (2.6%) | 2.06 | 0.88 |
| I am not aware of all the project stakeholders we are working with | 8 (2.6%) | 64 (20.7%) | 14 (4.5%) | 159 (51.5%) | 64 (20.7%) | 3.67 | 1.10 |
| Project reports are not disseminated to stakeholders of the project on a regular basis | 55 (17.8%) | 208 (67.3%) | 8 (2.6%) | 24 (7.8%) | 14 (4.5%) | 2.14 | 0.95 |
| **Composite Mean and Standard Deviation** | | | | | | **3.11** | **1.08** |

*Source: Field Data (2025)*

All 309 survey participants were asked to assess their level of agreement with certain statements about stakeholder involvement, communication, engagement forums, reporting, and decision-making, among other subjects. The assessment used a 5-point Likert scale, where 1 represented severe disagreement, 2 denoted disagreement, 3 signified neutrality, 4 indicated agreement, and 5 reflected strong agreement. A rating of 1 denoted total disagreement with the statement, whilst a rating of 5 indicated strong agreement. The findings are shown in Table 4.

The investigation revealed that most respondents confirmed the presence of regular opportunities for stakeholder interaction in the projects. The survey results indicated that 40.8% (126) of the respondents strongly agreed with this assertion. The mean rating was 4.05, surpassing the overall composite mean of 3.11, with a standard deviation of 1.065, compared to the overall standard deviation of 1.08.

The respondents also said that stakeholders contributed to the development of the organization/project ($\bar{x}$= 3.79, SD=.940). The survey data indicated that 72.2% (223) of the participants agreed with this assertion. The average rating was 3.79, above the overall composite mean of 3.11, with a standard deviation of 0.94, signifying less variability compared to the overall standard deviation of 1.08.

The initiatives regularly conveyed progress updates to stakeholders, especially the financial partners, as indicated by respondents who assigned a mean rating of 3.57 (SD=1.348). This surpassed the composite mean of 3.11 and the standard deviation of 1.08. A majority of respondents, namely 47.6% (147) of the participants in this study, indicated agreement. As a result, stakeholder engagement forums, contributions, and reports garnered mean scores ranging from 3.6 to 4.0, indicating that the significance of stakeholders in donor-funded programs is paramount.

However, most respondents said that they were unaware of all the parties engaged in the activities ($\bar{x}$=3.67, SD=1.099). The poll indicated that 51.5% (159) of the participants agreed with this conclusion. The average rating was 3.67, surpassing the overall composite mean of 3.11, with a standard deviation of 1.099, indicating higher variability in the responses relative to the general standard deviation of 1.08. This tendency proved hazardous, since efforts thrived more when they created a network of stakeholders and provided a platform for stakeholder engagement. This was a strategy for maintaining donor-funded initiatives.

The majority of respondents reported that the viewpoints of project stakeholders were integrated into the design of project activities and initiatives, that there was consistent communication with stakeholders concerning project execution, and that a defined communication and feedback framework was established within the organisation. The mean scores were 3.40 (SD=1.297), 3.37 (SD=0.974), and 3.19 (SD=1.343), respectively. All scores above the overall composite mean of 3.11 and the standard deviation of 1.099, indicating that most individuals in this research had positive reactions with less variability in their means relative to the general standard deviation of 1.08.

Furthermore, the preponderance of respondents confirmed that the organisations participated in stakeholder mapping and upheld detailed stakeholder lists ($\bar{x}$=2.39, SD=.983); stakeholders' perspectives and opinions were diligently incorporated into our programming ($\bar{x}$=2.06, SD=.879), and project reports were consistently distributed to project stakeholders ($\bar{x}$=2.14, SD=.952). It was noted that, although being project stakeholders, community members were mostly not considered significant stakeholders in the majority of projects ($\bar{x}$=2.61, SD=1.019). The bulk of replies were driven by convenience rather than a programming need. This was paradoxical, given that most programs flourished thanks to the backing of the communities in which they were implemented. Moreover, the community members understood their fundamental needs, rendering their participation in the effort crucial for success.

This measure had a composite mean of 3.11 and a standard deviation of 1.08, indicating that most respondents were generally positive towards claims about project stakeholders. This demonstrated that donor-funded initiatives acknowledged the significant influence of project stakeholders and that these efforts were profoundly affected by them.

Comparable results were reported by the key informants, as the majority of respondents affirmed that stakeholders were consulted. Nevertheless, other programs failed to sufficiently engage stakeholders, resulting in difficulties in reaching the people they were intended to serve. Several respondents noted that the projects occasionally exploited the community for financial profit.

**Table 5: Frequency of stakeholder engagement**

| Stakeholder Engagement | Frequency | Percent |
|---|---|---|
| Never | 28 | 9.1 |
| Rarely | 77 | 24.9 |
| Often | 188 | 60.8 |
| Very Often | 16 | 5.2 |
| **Total** | **309** | **100.0** |

*Source: Field Data (2025)*

The study aimed to evaluate the frequency of stakeholder involvement in the examined donor-funded projects. All 306 respondents participating in the survey provided a response to this. The results are depicted in Table 5 above. The findings indicated that stakeholders were frequently engaged in donor-funded initiatives in various capacities, as previously discussed. Sixty-point eight (60.8%, N=188) of the participants in this survey indicated this. The respondents emphasised that the programs relied significantly on stakeholders for implementation, despite the fact that the contributions of certain stakeholders were inconsequential to the execution of donor-funded projects. The average rating was 2.62 (SD = 0.722), indicating that stakeholders were frequently engaged in donor-funded projects.

The study findings correspond with Panday, Wang, & Spasova (2023), who found that participatory project planning in Khulna had a substantial impact on community engagement and participation, aligning with previous research. Spasova and Braungardt (2021) discovered a strong positive linear relationship between project scope and project performance, specifically about environmental conditions ($\beta=0.324$, $p=0.034$). Kanyi and James (2023) discovered a robust and favourable association between technical, management, and governance proficiency and the efficacy of initiatives sponsored by donors in Nairobi County.

## IV. CONCLUSION

The most active participants in programming were the project beneficiaries, project employees, and area local administration. The majority of stakeholders participated in community outreach initiatives, workshops, and meetings. Since stakeholders helped the organisation and its projects flourish, there were regular forums where they participated in the projects. The projects constantly reported on project progress to the project stakeholders, primarily the funding partners. Even if the project beneficiaries were unaware of all the stakeholders they were dealing with, the importance of stakeholders in donor-funded initiatives could not be overstated. The majority of respondents agreed that project stakeholders had an impact on donor-funded projects' performance because donor-funded initiatives appreciated their contributions.

Therefore, the study came to the conclusion that initiatives financed by donors welcomed stakeholder input and made every effort to integrate it into their programming. The initiatives heavily dependent on the stakeholders for project implementation, despite the suggestions/ input provided by certain stakeholders were of no significance in the implementation of donor-funded projects. The involvement of stakeholders was unavoidable, and programs had to do so if they were to achieve programmatic success and sustainability. According to the report, project managers working on donor-funded initiatives should increase stakeholder involvement in order to better understand community needs and get ready for transitions.

By implication, the findings of this study could be used to substantiate stakeholder theory since It was made clear that project stakeholders play a crucial role and that their absence could cause the project to fail. This fact has been mastered by donor-funded initiatives, which have made every effort to include stakeholders in their activities. Furthermore, by offering an evidence-based examination of the optimal scenario in donor-funded projects, this paper acts as a resource for academics looking to increase their understanding of M&E.

**Conflict of Interest**
The author declares that there is no conflict of interest concerning the publishing of this paper.